\documentclass[runningheads]{llncs}
\usepackage[T1]{fontenc}
\usepackage{graphicx}
\usepackage{multirow}
\begin{document}
\title{SHISRCNet: Super-resolution And Classification Network For Low-resolution Breast Cancer Histopathology Image}
%\author{Luyuan Xie, Cong Li, Zirui Wang, Xin Zhang, Boyan Chen, Qingni Shen, Zhonghai	Wu} 

%index{Luyuan, Xie}

\author{Luyuan Xie\inst{1}\and Cong Li\inst{1} \and Zirui Wang\inst{2}\and Xin Zhang\inst{1}\and Boyan Chen\inst{1}\and Qingni Shen\inst{1}\thanks{Corresponding author: qingnishen@ss.pku.edu.cn}\and Zhonghai Wu\inst{1}\thanks{Corresponding author: wuzh@pku.edu.cn}}
%index{Luyuan, Xie}
%index{Cong, Li}
%index{Zirui, Wang}
%index{Xin, Zhang}
%index{Boyan, Chen}
%index{Qingni, Shen}
%index{Zhonghai, Wu}

%\authorrunning{Anonymous et al.}
\institute{$^{1}$School of Software and Microelectronics, Peking University, Beijing, China \\ $^{2}$Tencent Cloud Media, Shenzhen, China  
\\ Our Code: https://github.com/xiely-123/SHISRCNet}
\maketitle              % typeset the header of the contribution
\begin{abstract}
The rapid identification and accurate diagnosis of breast cancer, known as the killer of women, have become greatly significant for those patients. Numerous breast cancer histopathological image classification methods have been proposed. But they still suffer from two problems. (1) These methods can only hand high-resolution (HR) images. However, the low-resolution (LR) images are often collected by the digital slide scanner with limited hardware conditions. Compared with HR images, LR images often lose some key features like texture, which deeply affects the accuracy of diagnosis. (2) The existing methods have fixed receptive fields, so they can not extract and fuse multi-scale features well for images with different magnification factors. To fill these gaps, we present a \textbf{S}ingle \textbf{H}istopathological \textbf{I}mage \textbf{S}uper-\textbf{R}esolution \textbf{C}lassification network (SHISRCNet), which consists of two modules: Super-Resolution (SR) and Classification (CF) modules. SR module reconstructs LR images into SR ones. CF module extracts and fuses the multi-scale features of SR images for classification. In the training stage, we introduce HR images into the CF module to enhance SHISRCNet's performance. Finally, through the joint training of these two modules, super-resolution and classified of LR images are integrated into our model. The experimental results demonstrate that the effects of our method are close to the SOTA methods with taking HR images as inputs.
%breast cancer histopathological
\end{abstract}
\section{Introduction}
Breast cancer is one of the high-mortality cancers among women in the 21st century.  Every year, 1.2 million women around the world suffer from breast cancer and about 0.5 million die of it \cite{bray2018global}.  Accurate identification of cancer types will make a correct assessment of the patient's risk and improve the chances of survival. However, the traditional analysis method is time-consuming, as it mainly depends on the experience and skills of the doctors. Therefore, it is essential to develop computer-aided diagnosis (CADx) for assisting doctors to realize the rapid detection and classification.
%Identifying a patient’s cancer type and benignity plays a crucial role in treatment.
%In recent years, a series of deep learning methods have been proposed to solve the breast histopathological image classification issue. Spanhol et al. \cite{breast_cancer_CNN} improved the AlexNet structure to classify breast histopathology images, which showed a significant improvement in recognition accuracy compared with the previous work using hand-extracted breast cancer image features for classification \cite{dataset,Akay2009SupportVM}. FEBkCapsNet \cite{wang2021automatic} proposed a deep feature fusion and enhanced routing framework, combining CNN for highlighting semantics and capsule network that focusing the detailed position information. DSoPN \cite{li2020breast} presented a robust global covariance pooling module based on matrix power normalization to explore second-order statistics of deep features. SSCA \cite{xu2022selective} utilized  feature pyramid network (FPN) and attention mechanism to extract discriminative features from complex backgrounds at different magnification factors.

Due to being collected by various devices, the resolution of histopathological images extracted may not always be high. Low-resolution (LR) images lack of lots of details, which will have an important impact on doctors' diagnosis. Considering the improvement of histopathological images’ acquisition equipment will cost lots of money while significantly increasing patients’ expense of detection. The super-resolution (SR) algorithms that improve the resolution of LR images at a small cost can be a practical solution to assist doctors in diagnosis. At present, most single super-resolution methods only have fixed receptive fields \cite{dong2014learning,lai2017deep,shahidi2021breast,ledig2017photo}. These models cannot capture multi-scale features and do not solve the problems caused by LR in various magnification factors well. MRC-Net \cite{chen2021super} adopted LSTM \cite{hochreiter1997long} and Multi-scale Refined Context to improve the effect of reconstructing histopathological images. It considered the problem of multi-scale, but only fused two scales features. This limits its performance in the scenarios with various magnification factors. Therefore, designing an appropriate feature extraction block for SR of the histopathological images is still a challenging task.

In recent years, a series of deep learning methods have been proposed to solve the breast cancer histopathological image classification issue by the high-resolution (HR) histopathological images. \cite{breast_cancer_CNN,wang2021automatic,li2020breast} improved the specific model structure to classify breast histopathology images, which showed a significant improvement in recognition accuracy compared with the previous works \cite{dataset,Akay2009SupportVM}. SSCA \cite{xu2022selective} considered the problem of multi-scale feature extraction which utilized feature pyramid network (FPN) \cite{lin2017feature} and attention mechanism to extract discriminative features from complex backgrounds. However, it only concatenates multi-scale features and does not consider the problem of feature fusion.  So it is still worth to explore the potential of extraction and fusion of multi-scale features for breast images classification.
%using hand-extracted breast cancer image features

%The above methods have achieved impressive performance on HR histopathological images. However, 

To tackle the problem of LR breast cancer histopathological images reconstruction and diagnosis, we propose the \textbf{S}ingle \textbf{H}istopathological \textbf{I}mage \textbf{S}uper-\textbf{R}esolution \textbf{C}lassification network (SHISRCNet) integrating Super-Resolution (SR) and Classification (CF) modules. The main contributions of this paper can be described as follows:

(1) In the SR module, we design a new block called Multi-Features Extraction block (MFEblock) as the backbone. MFEblock adopts multi-scale receptive fields to obtain multi-scale features. In order to better fuse multi-scale features,  a new fusion method named multi-scale selective fusion (MSF) is used for multi-scale features. These make MFEblock reconstruct LR images into SR images well.
\begin{figure*}[htb]
\setlength{\belowcaptionskip}{-0.3cm}
\centering
\includegraphics[width=0.95\textwidth]{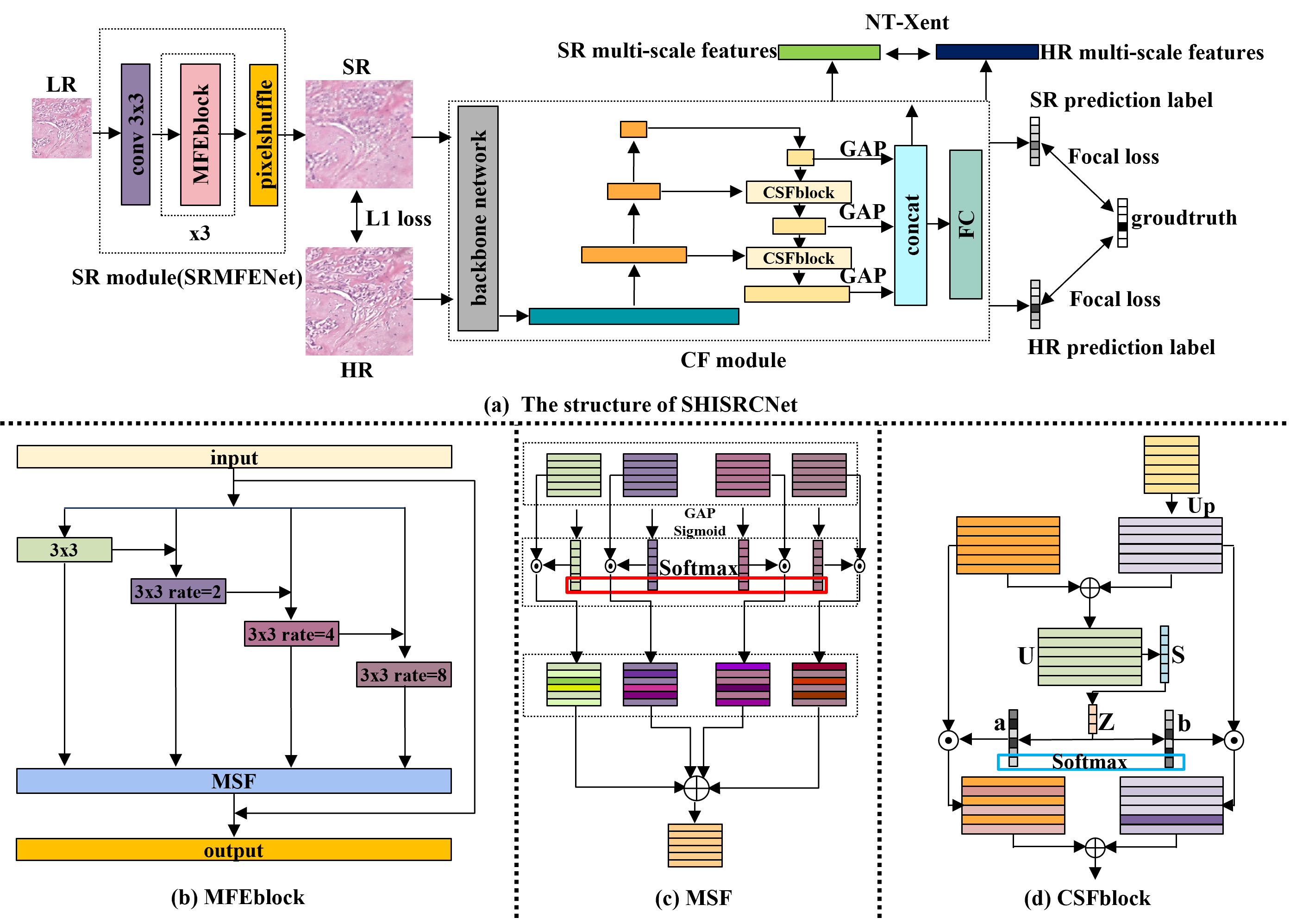}
\caption{ The structure of the SHISRCNet.}
\label{fig:traditional}
\end{figure*}

(2) The CF module completes the task of image classification by utilizing the SR images. Like SR module, it also needs to extract multi-scale features. The difference is that the CF module can use the method of downsampling to capture multi-scale features. So we combine the multi-scale receptive fields (SKNet) \cite{li2019selective} with the feature pyramid network (FPN) to achieve the feature extraction of this module. In FPN, we design a cross-scale selective fusion block (CSFblock) to fuse features of different scales.

(3) Through the joint training of these two designed modules, the super-resolution and classification of low-resolution histopathological images are integrated into our model. For improving the performance of CF module and reducing the error caused by the reconstructed SR images, we introduce HR images to CF module in the training stage. The experimental results demonstrate that the effects of our method are close to those of SOTA methods that take HR breast cancer histopathological images as inputs.
%\vspace{-0.1cm}
%\vspace{-0.5cm} 

\section{Methods}
%\vspace{-0.05cm}
This section describes the proposed SHISRCNet. The overall pipeline of the proposed network is shown in Fig. 1(a). It composes two modules: SR module and CF module. The SR module reconstructs the LR image into the SR image. The CF module utilize the reconstructed SR images to diagnose histopathological images. In the training stage, we introduce HR images to improve the performance of CF module and alleviate the error caused by SR images.

%As mentioned in Section \ref{sec:intro}.
%\vspace{-0.1cm} 
\subsection{Super-Resolution module}
%While collecting of histopathological images, images of the same patient at different magnification factors (e.g., 40x, 100x.etc) \cite{breast_cancer_CNN} are often collected. In addition, the images form the old equipments also face low-resolution (e.g., x2↓, x4↓, x8↓.etc) problems.
 To better extract and fuse multi-scale features for super-resolution, we propose a new SR network, called SRMFENet. Like SRResNet \cite{ledig2017photo}, SRMFENet takes a single low-resolution image as input and uses the pixelshuffle layer to get the restructured image. The difference between SRMFENet and SRResNet is that a Multi-Features Extraction block (MFEblock) is proposed to extract and fuse multi-scale histopathological images’ features. The structure of the MFEblock is shown in Fig. 1(b). The input features $X$ capture multi-scale features $Y_i$ through four 3$\times$3 atrous convolutions \cite{chen2017deeplab} with different rates:
%after adopting the convolution layers to draw features from the images. 

\[{Y_i} = \left\{ {\begin{array}{*{20}{c}}
{Cov3 \times {3_{rate = 1}}(X) }\\
{Cov3  \times {3_{rate = 2(i-1)}}(X + {Y_{i - 1}})}
\end{array}\begin{array}{*{20}{c}}
{i = 1}\\
{\quad 1 < i <= n }
\end{array}} \right. {\rm{ }}\]
where $n$ is the number of atrous convolutions and is set to 4 by the experiments. This design not only preserves the depth of the network, but also increases the width of the network. It is beneficial for the network to extract shallow local texture information and global semantic information. After the feature extraction phase, a new fusion method named MSF fuses all of different scale features $Y_i$. In the end, the input features $X$ are added with the fused features. The details of MSF show in the Fig. 1(c). Firstly, we conduct Global Average Pooling (GAP) \cite{lin2013network} on the multi-scale features to obtain their average channel-wise weights. Then using Sigmoid activation function to map weight to 0 to 1. Next softmax operation normalizes the same position of the obtained multi-scale average channel-wise weights. Finally, the features are multiplied by the corresponding normalized weights and the processed features are added together to generate the new multi-scale features. MFEblock is very applicable to process histopathological images of different magnification factors, as it employs convolution and attention operations to capture local and global image context information and fuse them well.

%Same as SR module, CF module also needs to extract multi-scale features. The difference is that
\subsection{Classification module}
 The task of the CF module is to classify the reconstructed SR images.  It can use the method of downsampling to capture multi-scale features. So we combine the multi-scale receptive fields (SKNet as backbone network) with the FPN (a downsampling method) to achieve the feature extraction of this module. In Fig. 1(a), the multi-sacle features extracted by SKNet are the input of FPN. We propose a new fusion method, called cross-scale selective fusion block (CSFblock) to effectively fuse high-resolution and low-resolution features in FPN. After the fused features are processed by GAP, they are aggregated into a new multi-scale feature by concatenate operation. Finally, the aggregated multi-scale features are classified through the fully connected (FC) layer. The structure of CSFblock is shown in Fig. 1(d). The inputs of CSFblock are two-way inputs which are the high-resolution features  $X_{h} \in W$×$H$×$C$ and the low-resolution features $X_{l} \in W_{1}$×$H_{1}$×$C$. In CSFblock, the upsampling operations are performed on the low-resolution features $X_{l}$ to realize consistency with $X_{h}$ dimension. $X_{h}$ and restructured $X_{l}$ are fused via an element-wise summation:
\[U{\rm{ }} = {\rm{ }}{X_h}{\rm{ }} + {\rm{ }}Up({X_l})\]
Then, using GAP along the channel dimension to get the global information $S$. A FC layer generates a compact feature vector $Z$ which guides the feature selection procedure. And $Z$ is reconstructed into two weight vectors $a$, $b$ of the same dimension as $S$ through two FC layers, which can be defined as:
\[Z{\rm{ }} = {\rm{ }}\delta ({W_c}S), \quad a{\rm{ }} = {\rm{ }}{W_a}Z,\quad b{\rm{ }} = {\rm{ }}{W_b}Z\]
Where $\delta$ denotes ReLU and $W_a$, $W_b$, $W_c$, means the weight of the FC layers. Specifically, a softmax operator is applied on $a$ and $b$ ’s channel-wise digits:
\[a[i]{\rm{ }} = {\rm{ }}\frac{{{e^{a[i]}}}}{{{e^{a[i]}} + {e^{b[i]}}}},\quad {\rm{ }}b[i]{\rm{ }} = {\rm{ }}\frac{{{e^{b[i]}}}}{{{e^{a[i]}} + {e^{b[i]}}}},\quad i \in C\]
The fused feature map $F$ is obtained through the attention weights on multi-scale features:
\[F[i]{\rm{ }} = {\rm{ }}a[i] \times {X_{h}}[i] + b[i] \times {Up({X_l})}[i],\quad i \in C\]
%Traditional features fusion methods usually adopt concatenate, add, and multiplication operations for different features directly. These methods are not effective for complex feature fusion tasks \cite{xu2022selective,li2019siamese,li2020bearing}. In the Classification module, we propose a new feature extraction and fusion method called Selective Feature block (SFblock) to generate discriminative features for classification. It uses convolution layers to extract details of high-resolution images and fuses multi-scale features from low-resolution images. SKNet \cite{li2019selective} selects the convolution kernels according to the size of the extraction target (Fig2(a)), so it can adjust the receptive fields of convolution according to the magnification factors of the histopathological image. In contrast with SKblock, the input of SFblock is two-way input X$_{HR}$ and X$_{SR}$, which are high-resolution features and the low-resolution features extracted by the SR module. To fuse X$_{HR}$ and X$_{SR}$  of different feature sizes, we designed SFblock-A, SFblock-B, and SFblock-C to adjust the input X$_{SR}$ feature size. As shown in Fig2 (c), (d), we use deconvolution and maxpooling respectively to upsample and downsample X$_{SR}$ to keep it the same as X$_{HR}$ in dimensions. No operation when X$_{HR}$ and X$_{SR}$ have the same dimensions (Fig2 (b)).

\subsection{Loss Function}
The SR module and the CF module exploit different loss functions for training. In the SR module, $L1$ Loss is used for super-resolution. In the CF module, we introduce HR images to CF module in the training stage for improving the performance of CF module and reducing the error caused by the reconstructed SR images. We use $Focal$ Loss \cite{lin2017focal} to alleviate the class imbalanced data problem of the HR and SR images' classification. Inspired by the contrastive learning algorithm SimCLR \cite{chen2020asimple}, the HR and SR of the
same images are similar to two different views. So the $NT-Xent$ loss \cite{sohn2016improved} is adopted to calculate similarity between SR multi-scale features and HR multi-scale features for CF module's robustness. The total loss function can be expressed as: 
\[{L_{total}}{\rm{ }} = {\rm{ }}{\lambda _1}{L_{{L1}}} + {\rm{ }}{\lambda _2}{L_{FL}} + {\rm{ }}{\lambda _3}{L_{NT-Xent}}{\rm{,\quad}}{\lambda _1} + {\lambda _2}+ {\lambda _3} = 1\]
where $\lambda_1$, $\lambda_2$ and $\lambda_3$ are the weights of $L1$ Loss, $Focal$ Loss and $NT-Xent$ Loss, respectively. In the inference stage, only SR images are taken as inputs by CF module. In our experiment, $\lambda_1$, $\lambda_2$ and  $\lambda_3$ are set to 0.6, 0.3 and 0.1, respectively. And the temperature parameter in $NT-Xent$ Loss is set to 0.5.
%\vspace{-0.1cm}
\section{Experiment}
%\vspace{-0.1cm}
\textbf{Dataset:} This work uses the breast cancer histopathological image database (BreaKHis) \footnote{https://web.inf.ufpr.br/vri/databases/breast-cancer-histopathological-database-breakhis/} \cite{dataset}. The images in the dataset have four magnification factors (40x, 100x, 200x, 400x) and eight breast cancer classes. This dataset includes four distinct histological types of benign breast tumors: adenosis (A), fibroadenoma (F), phyllodes tumor (PT), and tubular adenona (TA); and four malignant tumors (breast cancer): carcinoma (DC), lobular carcinoma (LC), mucinous carcinoma (MC) and papillary carcinoma (PC). The original dataset is randomly divided into training set and testing set for each magnification at a ratio of 7: 3 following previous work. 

\noindent
\textbf{Implementation Details:} For all experiments, we conduct 5-fold cross validation, and report the mean. We use LR histopathological images with size 48x48, 96x96, 192x192 as input for different single image SR tasks (x8, x4, x2) and set batch size to 8. For the corresponding LR and HR images in the training dataset, the same data augmentation is adopted, such as rotation, color jitter. The model is trained using the ADAM optimizer \cite{zhang2018improved} with the learning rate set to 1x$10^{-3}$. The learning rate is multiplied by 0.9 for every two epochs. We use SKNet-26 \cite{li2019selective} as the backbone network in the CF module. The total training epochs are 100.
%\vspace{-0.1cm} 
\section{Results and Discussion}
%\vspace{-0.1cm}
\subsection{The results of Super-Resolution and Classification}

\label{ssec:subhead}
%Table 1 shows the results of the super-resolution phase. We adopt Peak Signal to Noise Ratio (PSNR) and structural similarity index (SSIM) \cite{chen2021super} to evaluate the performance of the SR model. Our proposed SRMFENet (SR module) achieves better metrics than SRCNN, WA-SRGAN, EDSR and MRC-Net in three different scales. This proves that our MFEblocks can extract and fuse multi-scale features well. And the joint training of SRMFENet and CF module improves the super-resolution performance. Fig. 2 demonstrates that our model can recover more details with less blurring.

Table 1 shows the results of the super-resolution phase. We adopt Peak Signal to Noise Ratio (PSNR) and structural similarity index (SSIM) \cite{chen2021super} to evaluate the performance of the SR model. MRC-Net and our proposed SRMFENet (SR module) achieves better metrics than the other algorithms. This proves the effectiveness of multi-scale features extraction. Compared with  MRC-Net, our MFEblocks can extract and fuse multi-scale features well. And the joint training of SRMFENet and CF module improves the performance of super-resolution. Fig. 2 demonstrates that our model can recover more details with less blurring.
%%\vspace{-0.3cm} 
%\vspace{-0.2cm}
\begin{figure}[htb]
\centering
\includegraphics[width=0.9\textwidth]{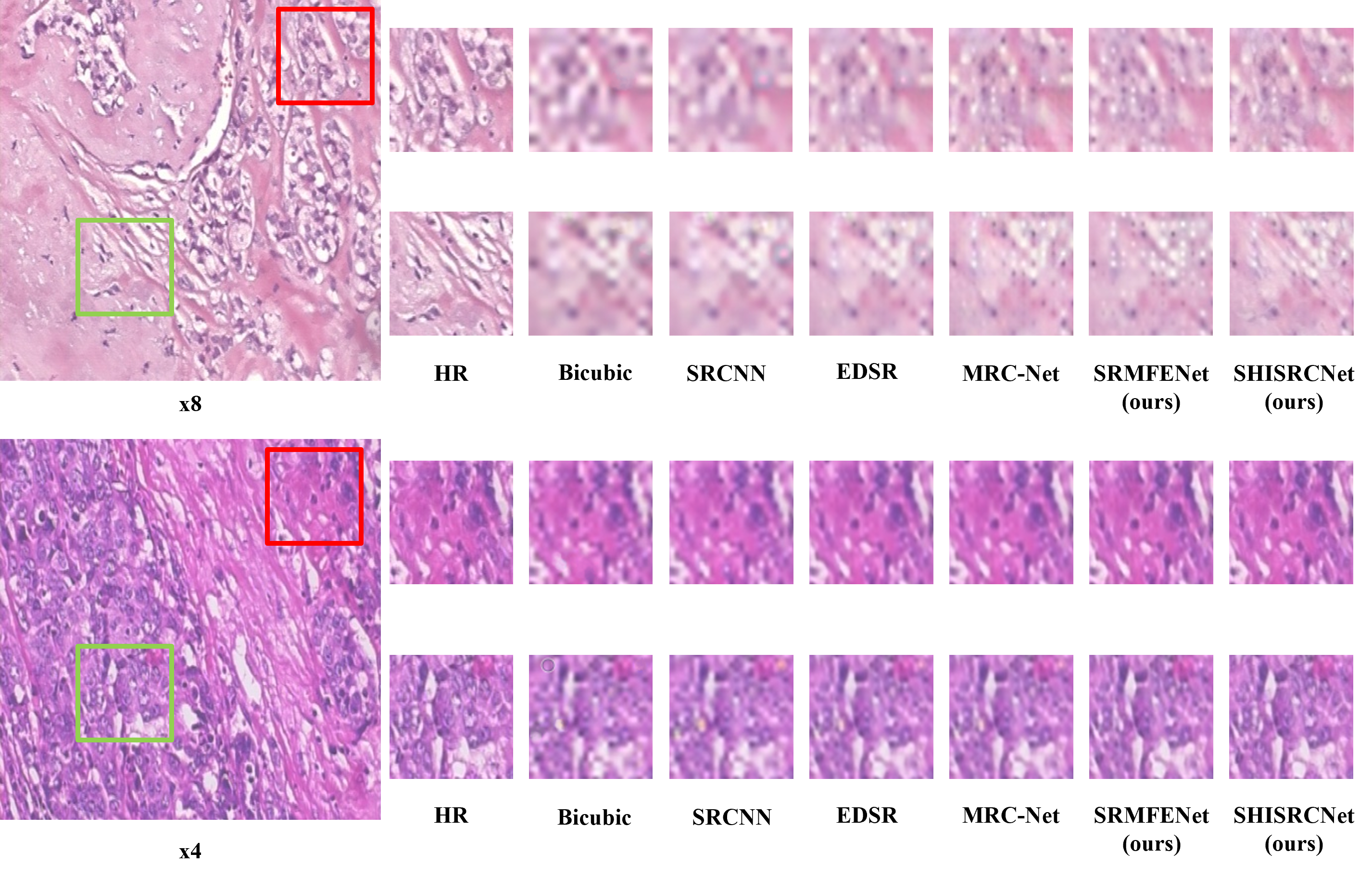}
\caption{ Qualitative Comparison with SR methods on breast cancer histopathological images x8 and x4.}
\label{fig:traditional}
\end{figure}

\begin{table}[htbp]
  \centering
  \caption{Average PSNR/SSIM for x8, x4, x2 SR.}
  
   \setlength{\tabcolsep}{2.3mm}{
   \scalebox{0.81}{
    \begin{tabular}{c|c|c|c|c|c|c}
     \hline
        \multirow{2}[0]{*}{Methods} & \multicolumn{2}{c|}{x8} & \multicolumn{2}{c|}{x4} & \multicolumn{2}{c}{x2} \\
        & PSNR(db)  & SSIM  & PSNR(db)  & SSIM  & PSNR(db)  & SSIM \\
     \hline
      \hline
    Bicubic & 20.75  & 0.4394  & 23.21  & 0.6305  & 26.60  & 0.9151 \\
    
    SRCNN & 21.12  & 0.4872  & 24.04  & 0.6634  & 28.36  & 0.8631  \\
    
    WA-SRGAN & 21.76  & 0.5141  & 26.20  & 0.7930  & 30.93  & 0.9351  \\
    
    EDSR  & 22.13  & 0.6063  & 26.22  & 0.8005  & 30.79  & 0.9325  \\
    
    MRC-Net & 22.72  & 0.6213  & 26.86  & 0.8222  & 31.73  & 0.9433  \\
    \hline
    SRMFENet (ours) & 23.44   & 0.6325   & 27.21  & 0.8370  & 33.96  & 0.9566  \\
    
    SHISRCNet (ours)  & \textbf{24.21} & \textbf{0.6814} & \textbf{27.97} & \textbf{0.8413} & \textbf{35.61} & \textbf{0.9680} \\
     \hline
    %\bottomrule
    \end{tabular}}}%
  \label{tab:addlabel}%
\end{table}%
%%\vspace{-1.0cm}

%\vspace{-0.1cm}
%%\vspace{-0.8cm} 

%\subsection{Comparison with State-of-the-art models}
We compare our introduced CF module with five state-of-the-art breast cancer histopathological image models and Diagnosis Network with MRC-Net \cite{chen2021super}, as shown in Table 2. The results illustrate that the CF module reaches the best performance in four different magnification factors. This indicates the effectiveness of our proposed combination of two multi-scale feature extraction methods.  SHISRCNet, which uses down sample to half resolution (x2↓) from HR images, outperforms the SSCA at 40x, 200x and 400x. And it gets results close to the CF module at all magnification factors. Meanwhile, compared with the Diagnosis Network that also uses LR images as input, SHISCNet has remarked performance advantages. Table 3 compares our results with the CF module using different resolution images. The performance of the CF module decreases significantly with the reduction of resolution. In contrast, SHISRCNet greatly improves the CF module performance of different scale low-resolution images.

%\vspace{-0.1cm}
\begin{table}[htbp]
  \centering
  \caption{Compare results with state-of-the-art on image level (* means that inputs are HR images, $^\#$ means that inputs are down sample to half resolution from HR images.).}
  \scalebox{0.8}{
   \setlength{\tabcolsep}{3.4mm}{
    \begin{tabular}{c|c|c|c|c|c}
    \hline
    \multirow{2}[0]{*}{Methods} & \multirow{2}[0]{*}{Years} & \multicolumn{4}{c}{ACC(\%)} \\
          &       & 40x   & 100x  & 200x  & 400x \\
      \hline
      \hline   
    AlexNet variant* \cite{breast_cancer_CNN} & 2016  & 85.6  & 83.5  & 82.7  & 80.7 \\
    Inception V3* \cite{benhammou2018first} & 2018  & 90.2  & 85.6  & 86.1  & 82.5 \\
    %SE-ResNet variant* \cite{gandomkar2018mudern} & 2019  & 94.43 & 94.45 & 92.27 & 91.15 \\
    DSoPN* \cite{li2020breast} & 2020  & 96    & 96.16 & 98.01 & 95.97 \\
    FE-BkCapsNet* \cite{wang2021automatic} & 2021  & 92.71 & 94.52 & 94.03 & 93.54 \\
    SSCA* \cite{xu2022selective}  & 2022  & 96.93 & 97.32 & 95.31 & 96.24 \\
   \hline
    CF module only* (ours) & —    & \textbf{97.82} & \textbf{97.78} & 
    \textbf{98.28} & \textbf{98.15} \\
    \hline
    Diagnosis Network with MRC-Net$^\#$ \cite{chen2021super} & 2021  & 94.43 & 94.45 & 94.73 & 93.92 \\
   \hline
    SHISRCNet (ours)$^\#$ & —    & \textbf{97.49} & \textbf{96.19} & \textbf{97.60} & \textbf{97.04} \\
    \hline
    \end{tabular}}}%
  \label{tab:addlabel}%
\end{table}%
%\vspace{-0.5cm} 

\label{ssec:subhead}

\begin{table}[htbp]
  \centering
  \caption{Comparison of accuracy under different scales on the image level.}
  \setlength{\tabcolsep}{5.5mm}{
  \scalebox{0.8}{
    \begin{tabular}{c|c|c|c|c|c}
    \hline
    \multirow{2}[0]{*}{resolution} & \multirow{2}[0]{*}{Model} & \multicolumn{4}{c}{ACC(\%)} \\
              &       & 40x   & 100x  & 200x  & 400x \\
              \hline
              \hline
        HR &  CF module only & \textbf{97.82} & \textbf{97.78} & \textbf{98.28} & \textbf{98.15} \\
        \hline
        \multirow{2}[0]{*}{x2↓ LR}    &  CF module only & 94.47  & 89.92  & 92.64  & 91.30  \\
              & SHISRCNet  & \textbf{97.49} & \textbf{96.19} & \textbf{97.60} & \textbf{97.04} \\
              \hline
        \multirow{2}[0]{*}{x4↓ LR}    &  CF module only & 90.32  & 87.71  & 88.61  & 86.89  \\
              & SHISRCNet  & \textbf{94.15} & \textbf{94.22} & \textbf{95.14} & \textbf{95.26} \\
              \hline
        \multirow{2}[0]{*}{x8↓ LR}    &  CF module only & 84.11  & 82.32  & 84.62  & 82.23  \\
              & SHISRCNet  & \textbf{91.47} & \textbf{92.43} & \textbf{92.98} & \textbf{92.78} \\
        
    \hline
    \end{tabular}}}%
  \label{tab:addlabel}%
\end{table}%
% Table generated by Excel2LaTeX from sheet 'Sheet2'
%\vspace{-0.1cm}  & 97.49 & 96.19 & 97.60 & 97.04

\subsection{Ablation study of the  SHISRCNet}
\label{ssec:subhead}
To verify the effectiveness of the proposed components in SHISRCNet, a comparison between SHISRCNet and its five components on x2↓ images is given in Table 3. (1) w/o MSF repalces MSF with concatenate operation and 1$\times$1 convolution. (2) w/o FPN + CSFblock means that only SKNet is used for feature extraction in the CF module. (3) w/o CSFblock, w/o HR images and w/o NT-Xent loss remove the corresponding operation, respectively. As shown in table 3, firstly, the performance of super-resolution in the SHISRCNet is significantly reduced when we remove MSF. It indicates the importance of MSF for multi-scale feature fusion in the SR module. Secondly, only SKNet is used to extract multi-scale features in the CF module, and the accuracy decreased significantly. This again proves the effectiveness of our proposed combination of two multi-scale feature extraction methods. Thirdly, compared with using FPN alone, the performance of  SHISRCNet is further improved by adding CSFblock to FPN. Finally, the introduction of HR images further promotes the performance of SHISRCNet. Because the training method of HR and SR images proposed by us helps to improve the generalization of the SHISRCNet.

%
% Table generated by Excel2LaTeX from sheet 'Sheet2'
\begin{table}[htbp]
  \centering
  \caption{Ablation study of SHISRCNet on x2↓ images.}
  \setlength{\tabcolsep}{3.4mm}{
  \scalebox{0.8}{
    \begin{tabular}{ccccccc}
    \hline
    \multirow{2}[0]{*}{} & \multirow{2}[0]{*}{PSNR(db)} & \multirow{2}[0]{*}{SSIM} & \multicolumn{4}{c}{ACC(\%)} \\
          &       &       & 40x   & 100x  & 200x  & 400x \\
    \hline
    \hline
     SHISRCNet & \textbf{35.61} & \textbf{0.9680} & \textbf{97.49} & \textbf{96.19} & \textbf{97.60} & \textbf{97.04} \\
    \hline
    w/o MSF & 33.13  & 0.9554  & 96.02  & 95.73  & 95.98  & 95.78  \\
    w/o FPN+CSFblock  & 34.41  & 0.9609  & 93.98  & 92.11  & 91.35  & 92.15  \\
    w/o CSFblock   & 34.57  & 0.9619  & 94.37  & 93.21  & 93.99  & 94.71  \\
    w/o HR images & 34.43  & 0.9611  & 93.13  & 94.28  & 93.86  & 94.17  \\
    w/o NT-Xent loss & 34.54  & 0.9623  & 95.53  & 95.20  & 95.01  & 95.36  \\
            \hline
    \end{tabular}}}%
  \label{tab:addlabel}%
\end{table}%
%\textbf{35.61} & \textbf{0.9680} \ & 97.49 & 96.19 & 97.60 & 97.04
%\vspace{-0.1cm}
\section{CONCLUSION}
\label{sec:majhead} 
This paper proposes SHISRCNet for the low-resolution breast cancer histopathological images’ super-resolution and classification problem. The SR module employs MFEblock to extract and fuse multi-scale features for reconstructing low-resolution histopathological images into high-resolution ones. The CF module adopts two different multi-scale features extraction methods to capture features for the breast cancer diagnosis. We introduce high-resolution images into the CF module in the training stage to improve SHISRCNet's robustness. Through the joint training of the two modules, the super-resolution and classification of the low-resolution histopathological images are integrated in one model. Our method’s results are close to the SOTA methods, which require using high-resolution breast cancer histopathological images instead of low-resolution ones.

\bibliography{reference}





% LNCS Homepage, \url{http://www.springer.com/lncs}. Last accessed 4
\end{document}